\documentclass[final,5p,times,twocolumn]{elsarticle}
\usepackage{amssymb}
\usepackage{color,soul}
\usepackage{dcolumn}% Align table columns on decimal point
\usepackage{graphicx}
\usepackage{lineno,hyperref}

\usepackage{amsmath}
\hypersetup{pdfborder={0 0 0}, pdfauthor={Author}, pdftitle={Title}, pdfkeywords={keywords}, pdfsubject={subject}, hidelinks}

\journal{Nuclear Inst. and Methods in Physics Research, A, }

\begin{document}

\begin{frontmatter}



\title{FCC-ee positron source from conventional to crystal-based}

\author[label1,label2]{Fahad Alharthi\corref{corr1}}

\author[label2]{Iryna Chaikovska}
\author[label2]{Robert Chehab}
\author[label2]{Viktor Mytrochenko} 
\author[label2]{Yuting Wang}

\author[label3]{Yongke Zhao} 

\author[label4]{Laura Bandiera} 
\author[label4]{Nicola Canale} 
\author[label4,label6]{Vincenzo Guidi} 
\author[label4]{Lorenzo Malagutti} 
\author[label4,label6]{Andrea Mazzolari} 
\author[label4,label6]{Riccardo Negrello} 
\author[label4]{Ginafranco Paternò\corref{corr2}} 
\author[label4,label6]{Marco Romagnoni} 
\author[label4]{Alexei Sytov}

\author[label5,label7]{Daniele Boccanfuso} 
\author[label5,label7]{Alberto Orso Maria Iorio} 

\author[label8]{Susanna Bertelli} 
\author[label8]{Mattia Soldani} 

\affiliation[label1]{organization={King Abdulaziz City for Science and Technology, KACST},
            city={Riyadh},
            country={Saudi Arabia}}
            
\affiliation[label2]{organization={Université Paris-Saclay, CNRS/IN2P3, IJCLab},
            city={Orsay},
            country={France}}

\affiliation[label3]{organization={CERN},
            city={Geneva},
            country={Switzerland}}
            
\affiliation[label4]{organization={INFN Sezione di Ferrara},
            city={Ferrara},
            country={Italy}}

\affiliation[label5]{organization={INFN Sezione di Napoli},
            city={Napoli},
            country={Italy}}

\affiliation[label6]{organization={Università degli Studi di Ferrara},
            city={Ferrara},
            country={Italy}}
            
\affiliation[label7]{organization={Università degli Studi di Napoli Federico II},
            city={Napoli},
            country={Italy}}

\affiliation[label8]{organization={INFN Laboratori Nazionali di Frascati},
            city={Frascati},
            country={Italy}}

\cortext[corr1]{Corresponding author; 
\textit{Email address:} 
\texttt{falharthi@kacst.gov.sa}}

\cortext[corr2]{Corresponding author; 
\textit{Email address:} \texttt{paterno@fe.infn.it}}
%% Abstract
\begin{abstract}

The high-luminosity requirement in future lepton colliders imposes a need for a high-intensity positron source. In the conventional scheme, positron beams are obtained by the conversion of bremsstrahlung photons into electron-positron pairs through the interaction between a high-energy electron beam and a high-Z amorphous target. One method to enhance the number of produced positrons is by boosting the incident electron beam power. However, the maximum heat load and thermo-mechanical stresses bearable by the target severely limit the beam power of the incident electrons. To overcome these limitations, an innovative approach using lattice coherent effects in oriented crystals appears promising.
This approach uses a single thick crystal that serves as a radiator and a converter. In this paper, we investigate the application of this scheme as an alternative to the conventional positron source at the Future Circular Collider (FCC-ee). Simulations were carried out from the positron production stage to the entrance of the damping ring to estimate the accepted positron yield. The results demonstrate the advantages of the crystal-based positron source: it requires thinner targets than the conventional scheme, resulting in a 14\% reduction in the deposited power while achieving a 10\% increase in accepted positron yield.
\end{abstract}

%% Keywords

\end{frontmatter}

% \linenumbers
\renewcommand{\figurename}{Fig.}

\section{Introduction}
\label{sec1}

The Future Circular Collider (FCC-ee) is a proposed high-luminosity particle collider that aims at significantly advancing our understanding of fundamental physics and shall serve as the successor to the Large Hadron Collider (LHC) at CERN. Designed to operate in several distinct modes, such as the $Z-pole$, $WW$ threshold, $t\bar{t}$ threshold, and $Higgs$ production, the FCC-ee will enable precise measurements of fundamental particles such as the Z boson, W boson, top quarks, and Higgs boson \cite{Benedikt2019}. Among these modes, $Z-pole$ is the most critical for the injector performance, as it requires a high stored current in the collider: 1.3~A \cite{Craievich2023}.  

The injector linac is a key element in achieving the required luminosity. It is foreseen to provide electron and positron beams for top-up injection in the two rings with a charge around 5~nC, at 100~Hz repetition rate and four bunches per linac pulse \cite{Craievich2023}.  
Within the injector complex, the positron ($e^{+}$) source is the most critical component. The $e^{+}$ source at FCC-ee relies on a conventional scheme, where a high-energy primary electron ($e^{-}$) beam impinges on an amorphous tungsten (W) target. This interaction generates bremsstrahlung radiation, which then converts into $e^{+} e^{-}$ pairs. Moreover, at much higher intensity, e.g., linear colliders, the $e^{+}$ source target faces several challenges. Notably, the energy deposition from the ionization losses in the target material can lead to excessive heating and potential damage to the target. The Peak Energy Deposition Density (PEDD) in the target can also cause localized energy deposits that create thermal gradients. These gradients can generate thermal stresses that may result in target failure \cite{chaikovska_positron_2022}.

An innovative approach based on the lattice coherent effects in an oriented crystal has been investigated to address these challenges. In 1989, Chehab et al. \cite{Chehab1989} proposed a $e^{+}$ source driven by coherent effects in oriented crystals. When a crystal is aligned along its strongest crystallographic axis, such as $\langle111\rangle$ for tungsten, it provides the highest mean electric field to the impinging charged particles. This enhances the interaction of high-energy electrons with the crystal lattice, leading to the production of soft photons in much greater abundance than those produced by conventional bremsstrahlung in an amorphous media. Since then, several experimental studies have confirmed the viability and relevance of this approach, demonstrating its potential to improve the $e^{+}$ production \cite{Artru, CHEHAB200241, Suwada}. The initial configuration splits the $e^{+}$ production into two stages: a thin crystal as a radiator followed by an amorphous target for $e^{+}$ production. 
Previous simulation studies showed that, the PEDD and the energy deposited in the amorphous is reduced \cite{artru_polarized_2008,SOLDANI2024168828}. However, as the gap between the crystal radiator and the amorphous target increases, the beam size grows, reducing the number of $e^{+}$ accepted by the downstream damping ring. Alternatively, we propose to use a single thick crystal, which acts as a radiator and a converter at the same time. 

An important parameter used to quantify the performance of the positron source is the accepted yield, i.e., a bunch charge of the $e^{+}$ accepted by the damping ring per the bunch charge of the primary $e^{-}$. Based on the simulation results, the FCC-ee $e^{+}$ source should be able to provide an accepted $e^{+}$ yield nine times higher than the current state-of-the-art hosted at KEK in Japan, for the SuperKEKB experiment (see Table \ref{table1}). This significant enhancement is mainly due to the use of a High-Temperature Superconduction (HTS) solenoid as a matching device and the large aperture of the Radio Frequency (RF) accelerating structures in the capture section.  

\begin{table}[h!]%% placement specifier
% \centering
\fontsize{7.2}{12}\selectfont

\caption{FCC-ee $e^{+}$ source parameters in comparison with the designed parameters of the SuperKEKB $e^{+}$ source.}\label{table1}
\begin{tabular}{p{5cm} c c}%% Table column specifiers
%% Tabular cells are separated by &
\hline
  Parameter & SuperKEKB \cite{chaikovska_positron_2022} & FCC-ee \\
\hline
  Primary $e^{-}$  energy [$\mathrm{GeV}$] & 3.5 & 2.86 \\
  Target thickness [$\mathrm{mm}$] & 14 & 15 \\
  Matching device, maximum field on the target [T] & 4.4 & 12 \\
  Maximum RF cavity aperture [$\mathrm{mm}$] & 30 & 60 \\
  Accepted $e^{+}$ yield at DR per GeV & 0.114 & 1.05 \\ 
\hline
\end{tabular}

\end{table} 

This paper provides an overview of the latest layout of the FCC-ee injector complex, primarily focusing on the $e^{+}$ source. Then, it presents and discusses the simulation results of the crystal-based $e^{+}$ source in application to the FCC-ee.

\section{FCC-ee injector and $e^{+}$ source}
\begin{figure}[h!]
\centering
\includegraphics[width=\columnwidth]{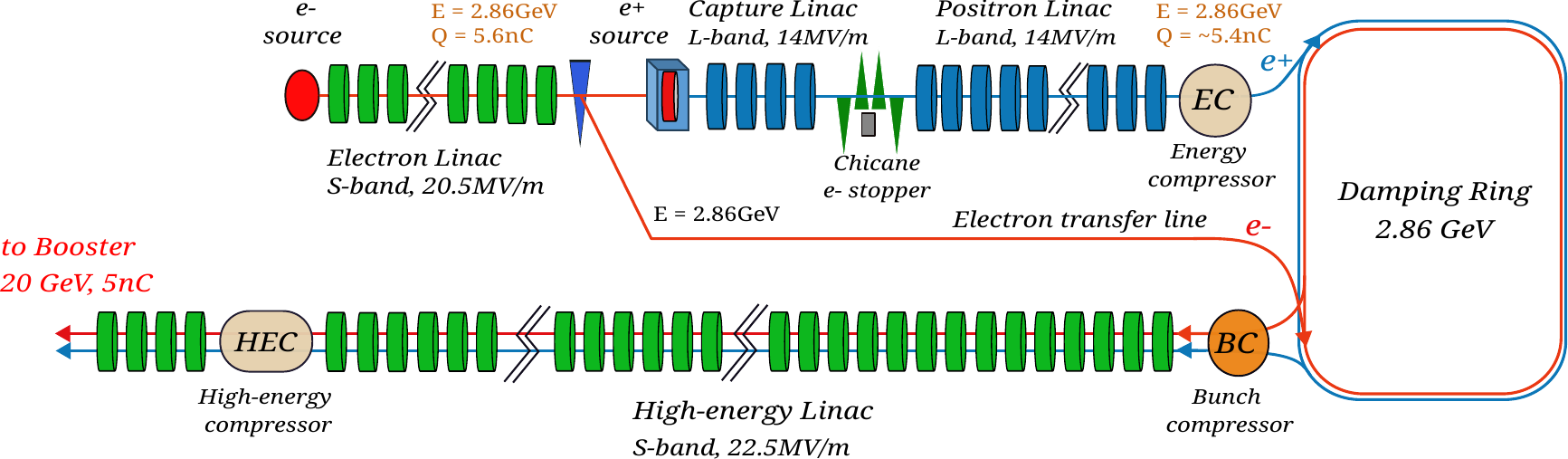}
\caption{The FCC-ee injector layout.}\label{fig1}
\end{figure}

The FCC-ee injector encompasses an $e^{-}$ source with a maximum bunch charge of 5.6~nC, an $e^{-}$ linac to accelerate the $e^{-}$ bunches up to 2.86~GeV, a $e^{+}$ source, a Damping Ring (DR) at 2.86~GeV for $e^{-}$/$e^{+}$ beams, and a High-energy linac to accelerate both particle species up to 20~GeV, which is the required energy at the booster injection\cite{zimmermann:ipac2024-wepr14}. The $e^{+}$ source is composed of a $e^{+}$ target, a capture section and a $e^{+}$ linac. A safety margin of 2.5 is applied for the whole FCC-ee $e^{+}$ source study\cite{chaikovska:ipac2023-mopl095}. The latest layout of the FCC-ee injector complex is illustrated in Fig.~\ref{fig1} 

In this context, the $e^{+}$ source is highly dependent on the parameters of the primary $e^{-}$, which are listed in Table \ref{tab:Table2}. 
\begin{table}[!h]
% \centering
\fontsize{7.2}{12}\selectfont

\caption{Parameters of the primary $e^{-}$ of the FCC-ee $e^{+}$ source}
    \label{tab:Table2}
    \begin{tabular}{p{7.3cm} c}
    \hline
         Beam parameters & Value \\
         \hline
         Energy [$\mathrm{GeV}$] & 2.86 \\
         Beam size ($\sigma_x$,$\sigma_y$) [$\mathrm{mm}$] & 1,1 \\ 
         Bunch length (RMS) [$\mathrm{mm}$] & 1 \\
         Energy spread (RMS) [\%]& 0.1 \\
         Normalized emittance [$\mathrm{mm} \cdot \mathrm{mrad}$] & 5.0 \\
         \hline
    \end{tabular}
\end{table}
The $e^{+}$ source target is made of tungsten and has a thickness of 15~mm. The thickness of the target has been optimized using the Geant4 toolkit \cite{agostinelli2003geant4,allison2006geant4,allison2016recent} to have the maximum number of $e^{+}$ generation. The sizeable angular spread of the generated $e^{+}$ (intrinsic feature due to the multiple scattering in the target) requires a focusing device, a so-called Adiabatic Matching Device (AMD) \cite{chehab_positron_1989}, which transforms the $e^{+}$ phase space to a smaller angular divergence and an acceptable transverse size, which fits within the dimensions of the following accelerating structures. In this setup, the target is placed inside the HTS solenoid, which presents a tapering magnetic field (AMD). The HTS solenoid is designed to generate a peak magnetic field of 15~T along the beam axis. This technology is the main factor for achieving a high accepted yield. An identical HTS solenoid is an integral part of the demonstrator for the FCC-ee $e^{+}$ source proof-of-principle at PSI (P\textsuperscript{3}) \cite{PhysRevAccelBeams.27.013401}. After the AMD, $e^{+}$ bunches are accelerated in six 3~m long large aperture L-band accelerating structures (2~GHz) with a gradient of 14~MV/m \cite{Hermann}. Each RF structure is surrounded by ten short solenoids, creating a solenoid channel of about 0.5~T to focus the $e^{+}$ bunches. The energy of the $e^{+}$ beam at the end of the capture linac is around 200~MeV. Note that a similar number of secondary $e^{-}$ are generated in the target and must be stopped. A magnetic chicane (four dipoles with a peak field of 0.2~T) is placed after the capture linac to separate the two charges. Therefore, the
secondary $e^{-}$ can be intercepted by a stopper at the middle of the chicane.. Since the $e^{+}$ linac and the DR are under extensive study and optimization, a simplified longitudinal tracking is used to boost the $e^{+}$ energy up to the DR energy. Then, the $e^{+}$ accepted yield is estimated by applying an energy-time window ($\pm$ 57.2~MeV, and $\pm$ 10~mm/$c$) on the longitudinal phase space. 

\begin{figure}[h!]
\centering
\includegraphics[width=\columnwidth]{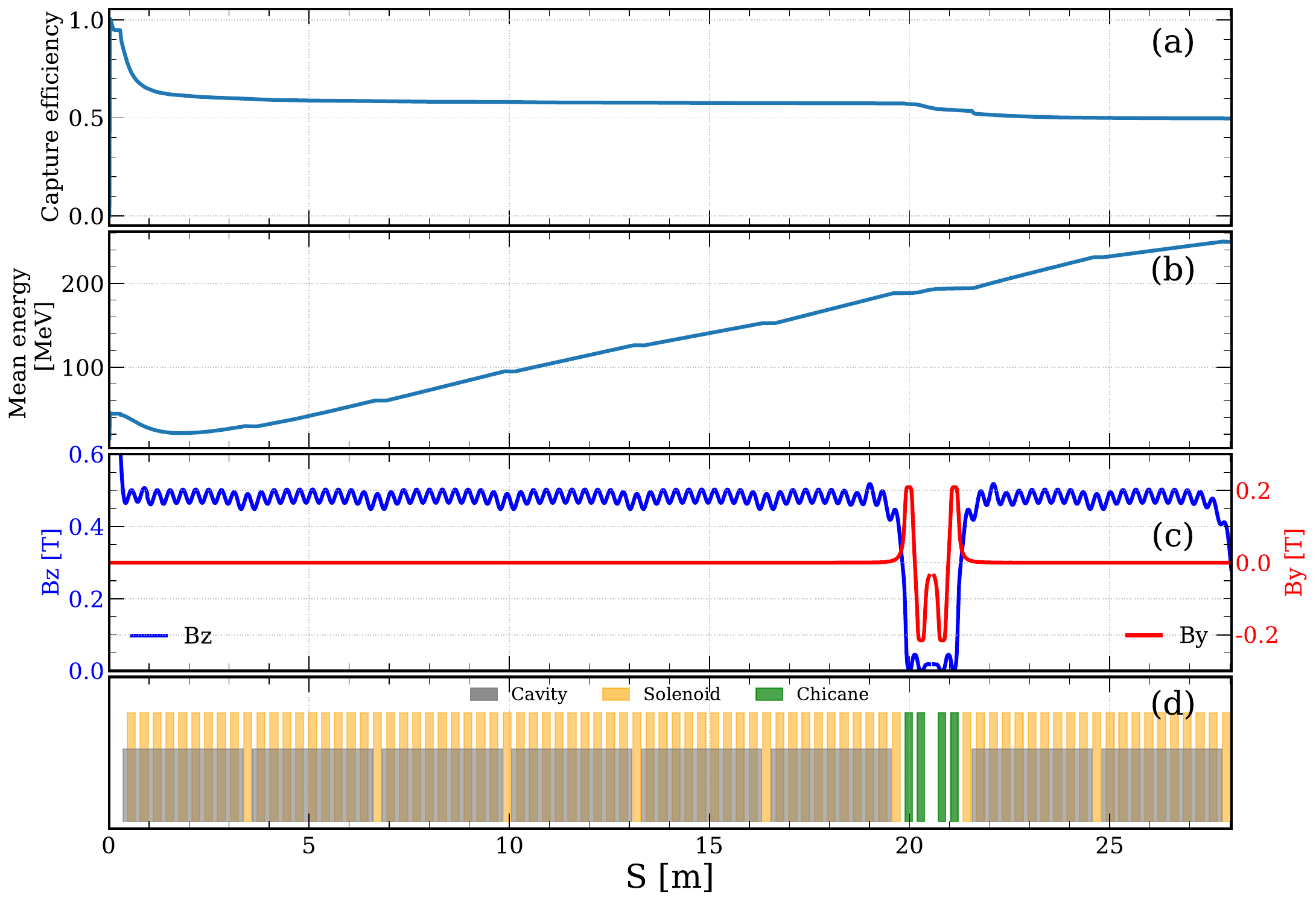}
\caption{$e^{+}$ tracking results along the longitudinal axis of the FCC-ee capture section, including capture efficiency (a), mean energy gain (b), magnetic field profiles (c), and schematics of the section's elements (d).}
\label{fig_tt}
\end{figure}

The $e^{+}$ beam tracking in the capture section is simulated using RF-Track \cite{RFT}. The RF phases of the accelerating structures in the capture linac are significantly influencing the accepted yield. X-opt machine learning optimization package \cite{Xopt}, based on the Bayesian optimization algorithm, is used to optimize the RF phases, thereby maximizing the accepted yield at the DR. Fig.~\ref{fig_tt} summarizes the particle tracking results in the capture section.

\begin{figure}[h!]
\centering
\includegraphics[width=\columnwidth]{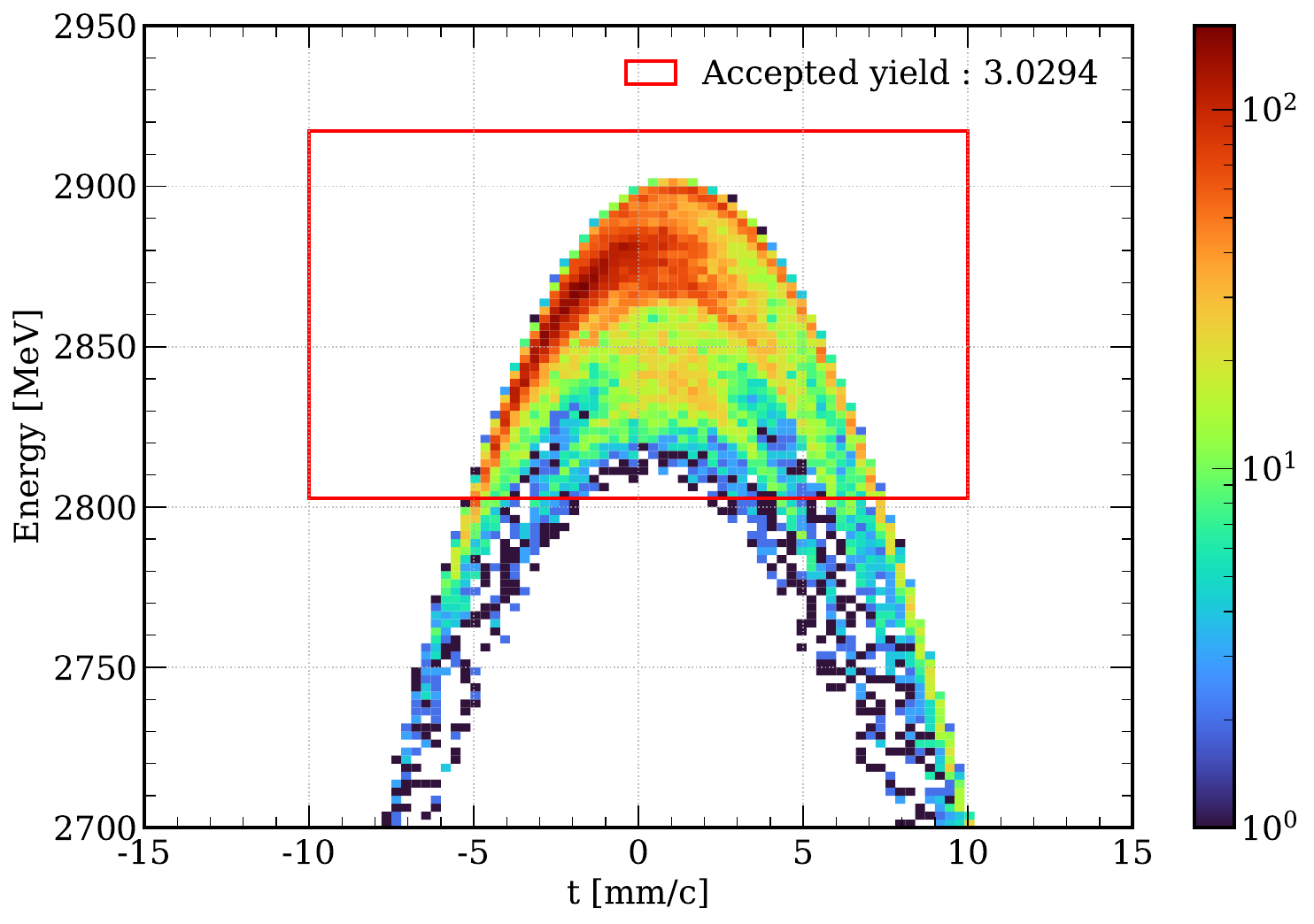}
\caption{$e^{+}$ longitudinal phase space at the end of the $e^{+}$ linac using the simplified longitudinal tracking. The red rectangle represents the energy-time window ($\pm$ 57.2~MeV, and $\pm$ 10~mm/$c$) centered at 2.86 $\mathrm{GeV}$ and the reference time set at 0.  }\label{fig_Long_phase}
\end{figure}

At the beginning of the capture section, the efficiency drops by 35\% due to the transverse acceptance of the RF accelerating structures. However, the $e^{+}$ beam losses remain negligible throughout the capture linac after this initial loss. Following the capture linac, losses in the chicane are around 9\%. With the current layout, an accepted yield of 3.03 is estimated after applying the cut window, as shown in Fig.~\ref{fig_Long_phase}. 

\begin{figure}[h!]
\centering
\includegraphics[width=\columnwidth]{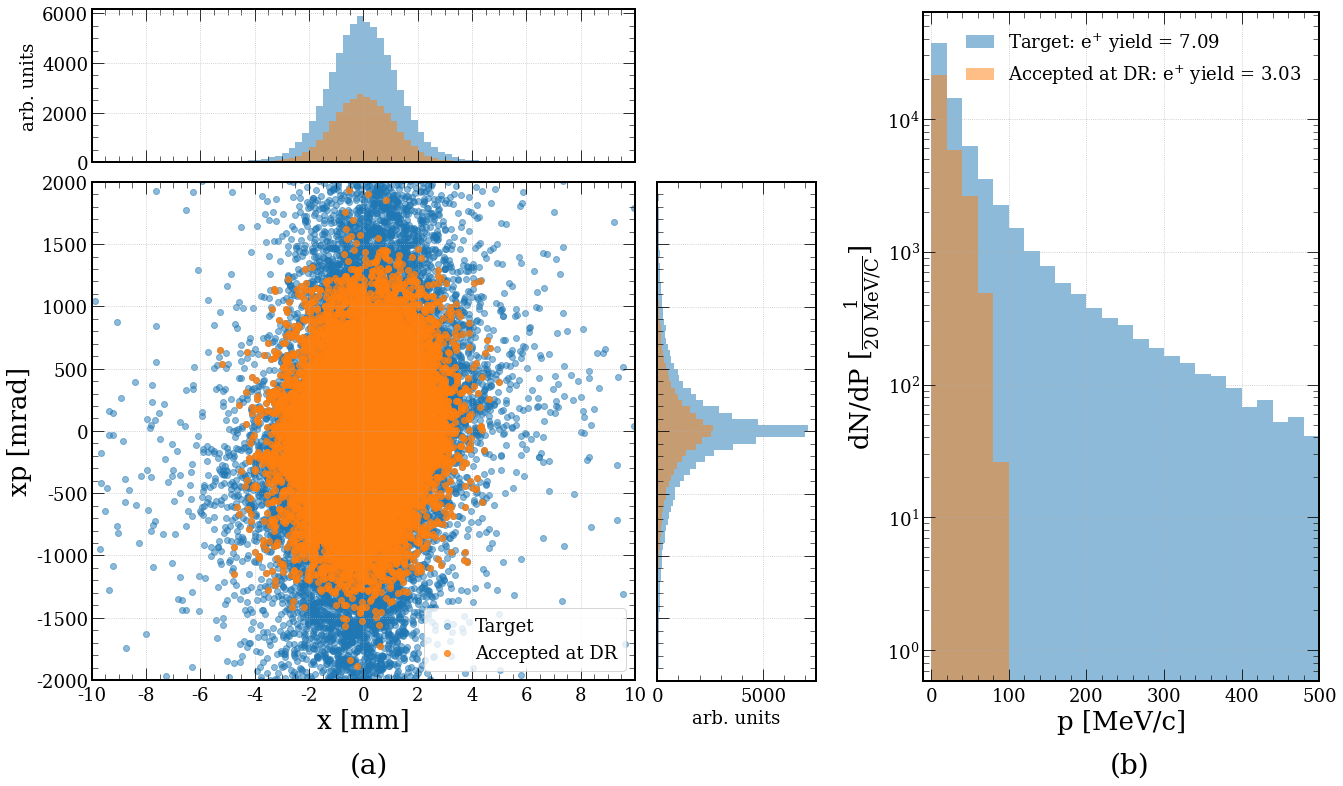}
\caption{$e^{+}$ distribution after the target, transverse phase space (a), and total momentum (b). Full distribution (blue), accepted by the DR (orange). }\label{fig_both}
\end{figure}

Fig.~\ref{fig_both} outlines the conditions for the accepted $e^{+}$ at the production stage. As expected, all the accepted $e^{+}$ have initial momentum below 100 $\mathrm{MeV/c}$; the primary factor in the yield enhancement. Meanwhile, the transverse size and the angular divergence play secondary roles. This motivates searching for alternative schemes, such as crystal-based $e^{+}$ sources, where the lattice coherent effects can further enhance the low energy $e^{+}$ production, thereby increasing the accepted yield.

\section{Crystal-based $e^{+}$ source modeling}
In order to optimize the crystal-based $e^{+}$ source for the FCC-ee, we developed a dedicated Geant4 application called \emph{PositronSource}. This application relies on the \texttt{G4ChannelingFastSimModel} \cite{Sytov2023, PhysRevAccelBeams.22.064601}, which is part of the Geant4 toolkit since version 11.2.0. The model simulates the coherent interactions of charged particles with the crystalline media, such as channeling and coherent pair production (since version 11.3.0). The classical trajectories of charged particles are simulated by integrating the equation of motion under the assumption of averaged atomic potential of the crystalline planes or axes \cite{sytov2015crystal}. Single and multiple scattering, as well as ionization, are evaluated at every simulation step. Also, photon emission is simulated using the \texttt{G4BaierKatkov} dedicated class via Monte Carlo integration of Baier-Katkov formula \cite{guidi2012radiation,bandiera2015radcharm++,bandiera2019compact,sytovPRAB2019,bandiera2022crystal,bandiera2024investigation}.
In order to simulate the lattice effects, a set of data related to the desired material and crystallographic orientation, such as the electric field, the electronic and nuclear densities, and the minimum ionization energy (at a given temperature) have to be calculated independently and then be imported into the Geant4 model\footnote{For more information, see the Geant4 documentation: \url{https://geant4-userdoc.web.cern.ch/UsersGuides/PhysicsReferenceManual/html/solidstate/channeling/channeling_fastsim.html}.}.
\emph{PositronSource} offers the flexibility to simulate a variety of crystal-based configurations and primary $e^{-}$ parameters. We aim to include this application among the official examples of Geant4 in the near future\footnote{In the meantime, \emph{PositronSource} is freely downloadable from the following link: \url{https://github.com/paternog/PositronSource}.}.

In our studies, we considered a single tungsten crystal aligned along its $\langle111\rangle$ crystallographic axis since this orientation provides the highest electrical field \cite{BaierKatkovStrakhovenko+1986+583+592}. The $e^{+}$ tracking was performed in the same FCC-ee capture section as used for the conventional scheme.

\begin{figure}[h!]
\centering
\includegraphics[width=\columnwidth]{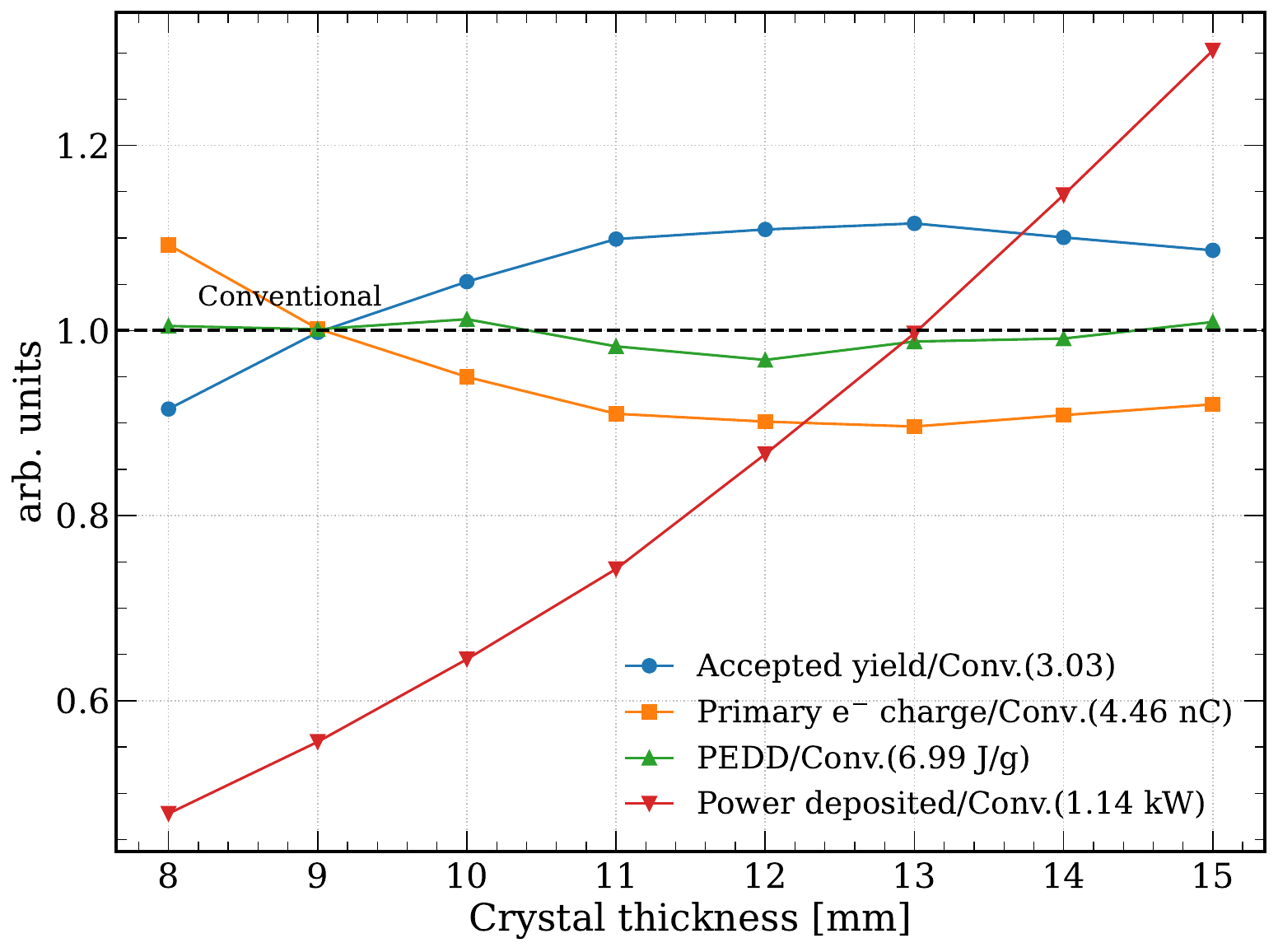}
\caption{Results of the simulation performed for the single tungsten crystal at room temperature. Note that all the crystal simulation results are normalized to those obtained for the conventional  $e^{+}$ source (dashed line represents the values on the legend).}\label{fig_opt}
\end{figure}

A scan of the crystal thickness was carried out to assess the performance of the $e^{+}$ source, as shown in Fig.~\ref{fig_opt}. The key parameters used to evaluate the crystal-based $e^{+}$ performance include the estimated accepted yield at the DR, the scaled primary $e^{-}$ bunch charge, the PEDD, and the total power deposited in the crystal. For comparison, we normalized all the values to the final results obtained for the conventional scheme. The simulation results show that the performance converges to a crystal thickness between 9~mm and 13~mm. For thicknesses above 9~mm, the accepted yield begins to surpass that of the conventional scheme, reaching a gain of about 10\%, accompanied by a significant reduction in energy deposition, which relaxes the target's cooling requirements.
Additionally, the PEDD remains similar to the values obtained for the conventional scheme.
This behavior can be explained by the fact that, as the crystal thickness increases, the production of photons and $e^{+}$$e^{-}$ pairs per primary electron is enhanced due to lattice coherent interactions. On the one hand, this causes an increase in energy density deposited per primary electron, with a peak value reached along the direction of the drive beam near the crystal exit. On the other hand, the improvement in the $e^{+}$ accepted yield allows us to reduce the drive beam current proportionally. Consequently, the total deposited power in the crystal remains below the value obtained for the conventional scheme up to a crystal thickness of 13~mm. Furthermore, the net effect on the PEDD does not vary significantly with the oriented crystal thickness.

The choice of a single tungsten crystal target with a thickness of 12~mm appears highly promising, as it provides an accepted yield slightly higher (+10\%) than that achieved with the conventional scheme while reducing the deposited power (-14\%). The final results are summarized in Table \ref{tab:Table3}.

\begin{table}[!h]
% \centering
\fontsize{7.2}{12}\selectfont

\caption{Summary of the FCC-ee $e^{+}$ source optimization results.}
    \label{tab:Table3}
    \begin{tabular}{p{4.8cm} c c}
    \hline
         Parameter & Conventional & Crystal-based \\
          \hline
         Target thickness [$\mathrm{mm}$] & 15 & 12  \\
         $e^{+}$ production rate  & 7.1 & 7.6 \\
         Accepted yield at the DR & 3.03 & 3.36 \\ 
         Primary $e^{-}$ bunch charge [$\mathrm{nC}$] & 4.46  & 4.0 \\
         Deposited power in the target [$\mathrm{kW}$]& 1.14 & 0.98 \\
         PEDD in the target [$\mathrm{J/g}$] & 6.99 & 6.76 \\
         \hline
    \end{tabular}
\end{table}

A possible limitation to the crystal-based $e^{+}$ source could come from the potential decrease of its performance due to the temperature increase caused by the energy deposit inside the crystal \cite{artru1997positron}. To assess this aspect, we evaluated the tungsten crystal $\langle111\rangle$ potential and the other required data at 600~K; then, we carried out the entire simulation, obtaining a negligible decrease (less than 1\%) in the accepted yield.

\begin{figure}[h!]
\centering
\includegraphics[width=\columnwidth]{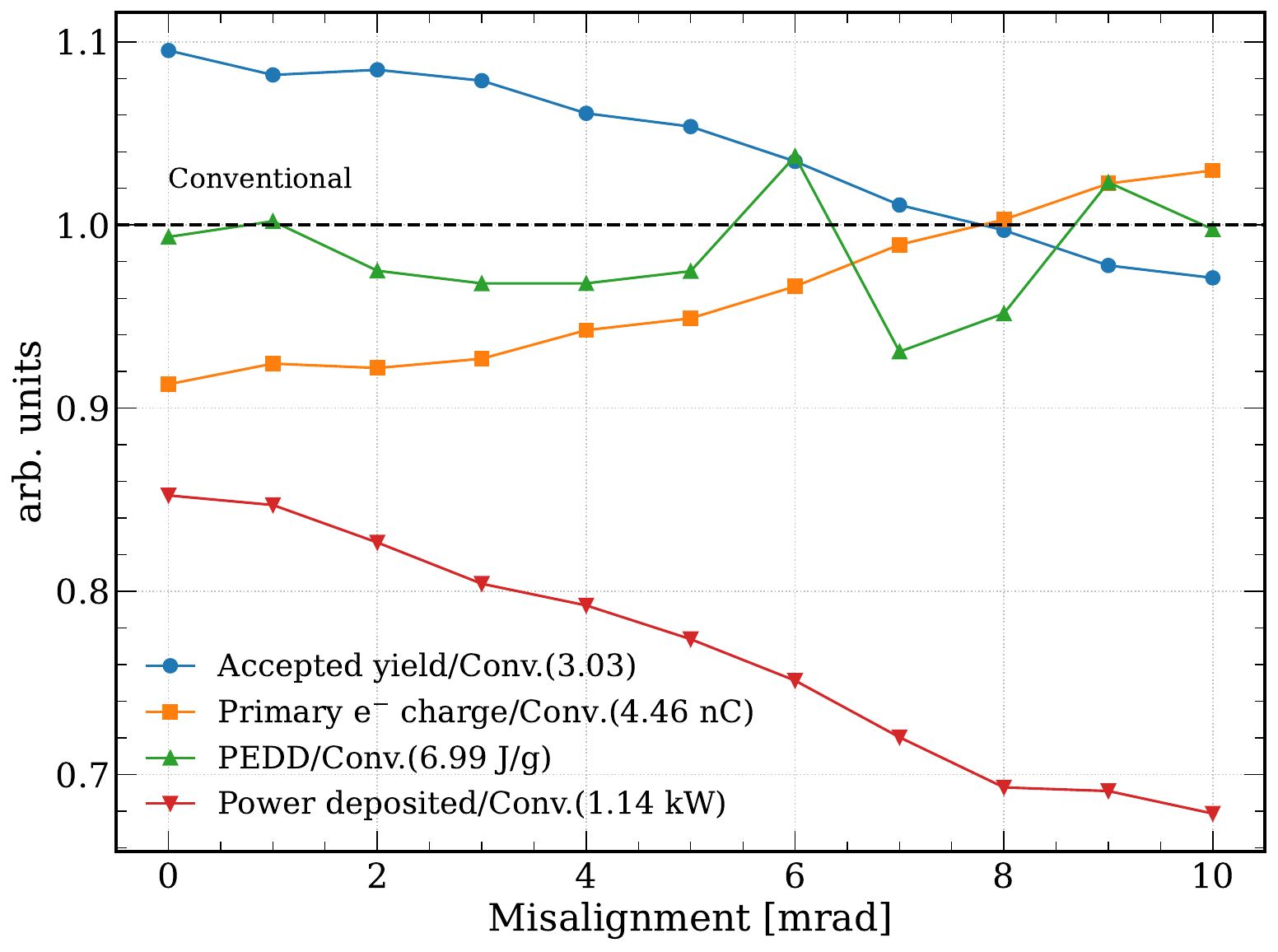}
\caption{Single tungsten crystal misalignment study at 600~K. Note that all the crystal simulation results are normalized to those obtained for the conventional $e^{+}$ source (dashed line represents the values on the legend).}\label{fig_mis}
\end{figure}

Another concern is that the crystal must be positioned inside the HTS solenoid without a remotely controlled goniometer, for which there is not enough space. Therefore, establishing the coherent effects must rely only on our ability to pre-align the crystal before placing it inside the HTS. The pre-alignment procedure, which we typically follow during our experiments involving oriented crystals, is based on a pre-characterization of the crystal surface angles with respect to the desired crystalline axis. This task is carried out by using an X-ray diffractometer \cite{romagnoni2022bent}. Then, the crystal sample is mounted on a cradle with six degrees of freedom during the test beam. Its surface is first aligned with the beam's nominal direction using a laser; then, the sample is rotated by pre-determined angles to align the desired crystalline axis with the beam. This procedure allows us to orient the crystal with a typical precision of 1 mrad \cite{bandiera2022crystal}. Furthermore, it is possible to improve this method further by substituting the laser with an autocollimator system, namely an optical instrument for non-contact measurement of angles \cite{sohn1998portable}, which works by projecting an laser onto a target mirror and automatically measuring the deflection of the reflected beam by means of an image sensor. The use of an autocollimator would guarantee superior precision in the alignment of the sample surface with the nominal beam direction.
We also conducted a set of dedicated simulations to evaluate the robustness of the crystal-based $e^{+}$ source performance in the presence of crystalline misalignment. The simulation involves rotating the crystal away from its axis and, at the same time, avoiding possible skew capture planes. The results of this study are illustrated in Fig.~\ref{fig_mis}. It is possible to notice a progressive decrease in the accepted positron yield accompanied by a decrease in the power deposited in the crystal, indicating a weakening of the coherent lattice interactions. Consequently, the drive beam current must be proportionally increased to compensate for this decrement. Nonetheless, the crystal-based $e^{+}$ source offers a performance advantage over the conventional scheme, even with misalignment of up to 8~mrad from the desired $\langle111\rangle$ axis. This level of misalignment is significantly larger than the typical precision achieved with our pre-alignment procedure. It is worth noting that this behavior was also observed experimentally in a thinner crystal with a 5.6~GeV electron beam \cite{bandiera2022crystal}.

% These results make us confident that this is an approach worth investigating further. In particular, it is important to study a suitable support system that allows pre-aligning the target, inserting it into the HTS, and preventing it from moving. Furthermore, a system must be studied to adequately cool the crystalline target and prevent its temperature from rising above 600~K.

\section{Conclusion and outlook}

This paper presents the latest FCC-ee $e^{+}$ source design based on the conventional scheme, including the capture section's full $e^{+}$ tracking results. The accepted yield by the DR is estimated to be around 3~$\frac{N_{e^{+}}}{N_{e^{-}}}$. Additionally, the conceptual design of the crystal-based $e^{+}$ source has been explored through various simulated options. The simulation results converge on using a single thick crystal that acts as a radiator and converter at the same time. This approach offers significant advantages, including a 14\% reduction in energy deposition and a 10\% increase in the accepted yield compared to the conventional scheme. 
On the other hand, crystal cooling and alignment inside the HTS cryostat are major challenges, even if a pre-alignment procedure could help mitigate this issue. To address these challenges, it is essential to study a suitable support system that allows pre-aligning the target, inserting it into the HTS solenoid, and preventing it from moving. Furthermore, a system must be studied to adequately cool the crystalline target and prevent its temperature from rising above 600~K.

\section*{Acknowledgements}
Work supported by ANR (Agence Nationale de la Recherche) Grant No: ANR-21-CE31-0007 and the European Union’s Horizon 2020 Research and Innovation programme under Grant Agreement No 101004730. This work was done under the auspices of CHART Collaboration (Swiss Accelerator Research and Technology). Also, we acknowledge financial support under the National Recovery and Resilience Plan (NRRP), Call for tender No. 104 published on 02.02.2022 by the Italian Ministry of University and Research (MUR), funded by the European Union – NextGenerationEU – Project Title : "Intense positron source Based On Oriented crySTals - e+BOOST" 2022Y87K7X– CUP I53D23001510006, and we acknowledge partial support of the INFN through the Geant4INFN project.  A. S. acknowledges the support from the H2020-MSCA-IF-Global TRILLION (G.A. 101032975).

\end{document}